\documentstyle[aps,pre,multicol]{revtex}

\def\bpsi{\mbox{\boldmath $\psi$}}
\def\bm{{\bf m}}
\def\bu{{\bf u}}
\def\bv{{\bf v}}
\def\bt{\bar{3}}
\def\bj{\bar{1}}

\begin{document}
\draft
\title{On the equivalence of the Ashkin-Teller and the four-state
Potts-glass models of neural networks}
\author{ D.~Boll\'e\thanks{E-mail:
           desire.bolle@fys.kuleuven.ac.be.}
           \thanks{Also at Interdisciplinair Centrum
            voor Neurale Netwerken, K.U.Leuven, Belgium.}
            and
         P.~Koz{\l}owski
          \footnotemark[2]
           \thanks{E-mail:
           piotr.kozlowski@fys.kuleuven.ac.be}
            }
\address{Instituut voor Theoretische Fysica,
            Katholieke Universiteit Leuven, \\ B-3001 Leuven, Belgium}

\maketitle
\begin{abstract}
\noindent
We show that for a particular choice of the coupling parameters
the Ashkin-Teller spin-glass neural network model with the Hebb learning
rule and one condensed pattern yields the same thermodynamic properties 
as the four-state
anisotropic Potts-glass neural network model. This equivalence is not
seen at the level of the Hamiltonians.
\end{abstract}

\begin{multicols}{2}
\narrowtext
It is well-known that the classical Ashkin-Teller (AT) model
is a generalization of the Ising, the four-state clock and the
four-state Potts models. This can be easily seen already at the level of the
Hamiltonian, especially when one rewrites the Hamiltonian of the
AT model \cite{AT}
using two Ising spins located at each site of the lattice interacting
via two- and four-spin couplings \cite{Fan}.

For spin-glass systems similar observations can be made (\cite{NS} and
references therein). The
AT spin-glass Hamiltonian contains as particular limits, for
certain bond realizations, both the four-state clock spin glass 
and the four-state Potts-glass Hamiltonians.

Concerning neural network models, the situation is more complicated.
It is straightforward to see at the level of the Hamiltonian
that for two, respectively one of the coupling strengths taken to be
zero, the AT neural network
model \cite{BK1,BK2} is equivalent to the Hopfield model \cite{Hop}
respectively the four-state clock neural network model \cite{Cook}.
On the contrary, the possible relation with the four-state Potts
neural network models existing in the literature \cite{Kanter,vz} is, at
first sight, unclear.
However, since we discovered in the study of the thermodynamic and
retrieval properties of the AT neural network
\cite{BK1,BK2} for  equal coupling
strengths some resemblance to the properties of the Potts-glass neural
network \cite{Kanter,BDH}, we expect that a relation with the
latter does exist.
To investigate this relation is the purpose of this brief report.

The AT neural network with the Hebb learning rule is described by the 
following infinite-range Hamiltonian
\begin{eqnarray}
   {\cal H}_{AT}=-\frac{1}{2N}  \sum_{\mu=1}^p  \sum_{(i,j)=1}^N
      \left[J_1 \xi_i^\mu \xi_j^\mu  s_i s_j
      + J_2  \eta_i^\mu \eta_j^\mu \sigma_i \sigma_j \right.
      \nonumber\\
       \left. + J_3  \xi_i^\mu \eta_i^\mu\xi_j^\mu\eta_j^\mu  s_i 
                           \sigma_i s_j \sigma_j\right]
           \label{at}
\end{eqnarray}
with the two types of Ising neurons $s_i,\sigma_i, i=1,
\ldots, N$ describing the state of the network. In this model storage
and retrieval of the  patterns $\{\xi_i^\mu\},\{\eta_i^\mu\},
\mu=1,\ldots,p$  are studied. The patterns
are randomly chosen configurations of the network. Based upon our
observations mentioned above we take equal coupling strengths
$J_1=J_2=J_3=1$ in the sequel.

Two Potts-glass neural networks with Hebb learning have been studied in
the literature. Considering four Potts-states the
first network \cite{Kanter} is described by the Hamiltonian
\begin{equation}
 {\cal H}_{K}=-\frac{1}{2N}  \sum_{\mu=1}^p  \sum_{(i,j)=1}^N
          \frac{1}{16}\left(\bu_i\cdot\bpsi_i^{\mu} \right)
         \left(\bu_j\cdot\bpsi_j^{\mu}\right) \, ,
         \label{kan}
\end{equation}
while the second one is given by \cite{vz}
\begin{equation}
  {\cal H}_{VZ}=-\frac{1}{2N}  \sum_{\mu=1}^p  \sum_{(i,j)=1}^N
               \frac{1}{16}\left(\bu_i\cdot\bu_j \right)
             \left(\bpsi_i^{\mu}\cdot\bpsi_j^{\mu} \right)
          \label{zip}  \, ,
\end{equation}
where $\bu_i$ and $\bpsi_i$ are state and pattern vectors taken from the
set of four-dimensional vectors $\bv = \{\bv^{(l)}\}$ with
components  $v^{(l)}_ k = 4\delta_{kl}-1$ for $l,k=1,2,3,4$.
The main difference between the two models is that in the first,
anisotropic model precisely one specific Potts state is favoured at each
site, while in the second, isotropic model the fact whether or not two
neurons are in the same state is important.

Two models can be equivalent at the level of the Hamiltonian or at the
level of the free energy. It is clear that for the Hamiltonians
of the models we have introduced above there are one state and $p$ pattern
variables associated with each of the $N$ sites of the network.
Thus the Hamiltonians can be written in the form
\begin{equation}
{\cal H}_{mod}=-\frac{1}{2N}\sum_{\mu=1}^{p}\sum_{(i,j)}H_{mod}
(C_{ij}^\mu)
\label{ham}
\end{equation}
where $mod$ denotes $AT$, $K$ or $VZ$.
The energy of the interaction between two sites is a sum over patterns of
$H_{mod}(C_{ij}^\mu)$ and depends on the state-pattern configuration
$C_{ij}^\mu$ of sites $i$ and $j$. Hence, it is enough to compare the
$H_{mod}(C_{ij}^\mu)$. In the case of four state models we are
considering here, all possible values of  $H_{mod}(C_{ij}^\mu)$  can be
written  in the form of a $16\times 16$ matrix ( 16 state-pattern
configurations for a given site).
For the sake of easy comparison we write down these matrices
explicitly\\
\begin{equation}
H_{K}=
\footnotesize
\left(\begin{array}[h]{cccccccccccccccc}
9&\bt&\bt&\bt&\bt&9&\bt&\bt&\bt&\bt&9&\bt&\bt&\bt&\bt&9\\
\bt&1&1&1&1&\bt&1&1&1&1&\bt&1&1&1&1&\bt\\
\bt&1&1&1&1&\bt&1&1&1&1&\bt&1&1&1&1&\bt\\
\bt&1&1&1&1&\bt&1&1&1&1&\bt&1&1&1&1&\bt\\
\bt&1&1&1&1&\bt&1&1&1&1&\bt&1&1&1&1&\bt\\
9&\bt&\bt&\bt&\bt&9&\bt&\bt&\bt&\bt&9&\bt&\bt&\bt&\bt&9\\
\bt&1&1&1&1&\bt&1&1&1&1&\bt&1&1&1&1&\bt\\
\bt&1&1&1&1&\bt&1&1&1&1&\bt&1&1&1&1&\bt\\
\bt&1&1&1&1&\bt&1&1&1&1&\bt&1&1&1&1&\bt\\
\bt&1&1&1&1&\bt&1&1&1&1&\bt&1&1&1&1&\bt\\
9&\bt&\bt&\bt&\bt&9&\bt&\bt&\bt&\bt&9&\bt&\bt&\bt&\bt&9\\
\bt&1&1&1&1&\bt&1&1&1&1&\bt&1&1&1&1&\bt\\
\bt&1&1&1&1&\bt&1&1&1&1&\bt&1&1&1&1&\bt\\
\bt&1&1&1&1&\bt&1&1&1&1&\bt&1&1&1&1&\bt\\
\bt&1&1&1&1&\bt&1&1&1&1&\bt&1&1&1&1&\bt\\
9&\bt&\bt&\bt&\bt&9&\bt&\bt&\bt&\bt&9&\bt&\bt&\bt&\bt&9
\end{array}
\right)
\end{equation}

\begin{equation}
H_{VZ}=
\footnotesize
\left(\begin{array}[h]{cccccccccccccccc}
9&\bt&\bt&\bt&\bt&1&1&1&\bt&1&1&1&\bt&1&1&1\\
\bt&9&\bt&\bt&1&\bt&1&1&1&\bt&1&1&1&\bt&1&1\\
\bt&\bt&9&\bt&1&1&\bt&1&1&1&\bt&1&1&1&\bt&1\\
\bt&\bt&\bt&9&1&1&1&\bt&1&1&1&\bt&1&1&1&\bt\\
\bt&1&1&1&9&\bt&\bt&\bt&\bt&1&1&1&\bt&1&1&1\\
1&\bt&1&1&\bt&9&\bt&\bt&1&\bt&1&1&1&\bt&1&1\\
1&1&\bt&1&\bt&\bt&9&\bt&1&1&\bt&1&1&1&\bt&1\\
1&1&1&\bt&\bt&\bt&\bt&9&1&1&1&\bt&1&1&1&\bt\\
\bt&1&1&1&\bt&1&1&1&9&\bt&\bt&\bt&\bt&1&1&1\\
1&\bt&1&1&1&\bt&1&1&\bt&9&\bt&\bt&1&\bt&1&1\\
1&1&\bt&1&1&1&\bt&1&\bt&\bt&9&\bt&1&1&\bt&1\\
1&1&1&\bt&1&1&1&\bt&\bt&\bt&\bt&9&1&1&1&\bt\\
\bt&1&1&1&\bt&1&1&1&\bt&1&1&1&9&\bt&\bt&\bt\\
1&\bt&1&1&1&\bt&1&1&1&\bt&1&1&\bt&9&\bt&\bt\\
1&1&\bt&1&1&1&\bt&1&1&1&\bt&1&\bt&\bt&9&\bt\\
1&1&1&\bt&1&1&1&\bt&1&1&1&\bt&\bt&\bt&\bt&9\\
\end{array}
\right)
\end{equation}

\begin{equation}
H_{AT}=
\footnotesize
\left(\begin{array}[h]{cccccccccccccccc}
3&\bj&\bj&\bj&\bj&3&\bj&\bj&\bj&\bj&3&\bj&\bj&\bj&\bj&3\\
\bj&3&\bj&\bj&3&\bj&\bj&\bj&\bj&\bj&\bj&3&\bj&\bj&3&\bj\\
\bj&\bj&3&\bj&\bj&\bj&\bj&3&3&\bj&\bj&\bj&\bj&3&\bj&\bj\\
\bj&\bj&\bj&3&\bj&\bj&3&\bj&\bj&3&\bj&\bj&3&\bj&\bj&\bj\\
\bj&3&\bj&\bj&3&\bj&\bj&\bj&\bj&\bj&\bj&3&\bj&\bj&3&\bj\\
3&\bj&\bj&\bj&\bj&3&\bj&\bj&\bj&\bj&3&\bj&\bj&\bj&\bj&3\\
\bj&\bj&\bj&3&\bj&\bj&3&\bj&\bj&3&\bj&\bj&3&\bj&\bj&\bj\\
\bj&\bj&3&\bj&\bj&\bj&\bj&3&3&\bj&\bj&\bj&\bj&3&\bj&\bj\\
\bj&\bj&3&\bj&\bj&\bj&\bj&3&3&\bj&\bj&\bj&\bj&3&\bj&\bj\\
\bj&\bj&\bj&3&\bj&\bj&3&\bj&\bj&3&\bj&\bj&3&\bj&\bj&\bj\\
3&\bj&\bj&\bj&\bj&3&\bj&\bj&\bj&\bj&3&\bj&\bj&\bj&\bj&3\\
\bj&3&\bj&\bj&3&\bj&\bj&\bj&\bj&\bj&\bj&3&\bj&\bj&3&\bj\\
\bj&\bj&\bj&3&\bj&\bj&3&\bj&\bj&3&\bj&\bj&3&\bj&\bj&\bj\\
\bj&\bj&3&\bj&\bj&\bj&\bj&3&3&\bj&\bj&\bj&\bj&3&\bj&\bj\\
\bj&3&\bj&\bj&3&\bj&\bj&\bj&\bj&\bj&\bj&3&\bj&\bj&3&\bj\\
3&\bj&\bj&\bj&\bj&3&\bj&\bj&\bj&\bj&3&\bj&\bj&\bj&\bj&3
\end{array}
\right)
\end{equation}
where we have used the standard notation $\bar{c}=-c$.

One can see that $H_{AT}$ differs from $H_{K}$ and $H_{VZ}$ by the
number of energy levels and by their position in the matrix. We have not
found a transformation of the Hamiltonians removing this difference.
Therefore, we conclude that at the level of the Hamiltonian the AT neural 
network
is not equivalent to any of the two four-state Potts models.

In order to find out whether a possible equivalence exists on the level
of the free energies we start from the model with Hamiltonian
${\cal H}_{K}$ because the matrix $H_K$ has the same global symmetry as
$H_{AT}$. We note that the
symmetry of $4 \times 4$ blocks in $H_{K}$ and $H_{AT}$ is different. It
is a consequence of the fact that ${\cal H}_{AT}$ is invariant under 
inversion of all the spins, while ${\cal H}_{K}$ is not
invariant under any permutation of the state variables. The Hamiltonian
${\cal H}_{VZ}$ on the contrary is completely invariant under any
permutation of those variables.

As remarked in \cite{Kanter} $H_{K}$ can be rewritten using
two different types of Ising spins
\begin{eqnarray}
  H_{K}(C_{ij}^\mu)&=&(s_i\xi_i^\mu + \sigma_i\eta_i^\mu+
                s_i\xi_i^\mu \sigma_i\eta_i^\mu)
        \nonumber\\
              &\times&(s_j\xi_j^\mu + \sigma_j\eta_j^\mu+
                      s_j\xi_j^\mu \sigma_j\eta_j^\mu)
            \label{split}
\end{eqnarray}
Applying the usual replica method \cite{mpv} to calculate the quenched
average over the patterns, chosen to be independent identically
distributed random variables taking the values $+1$ and $-1$ with equal
probability, the free energy density can be
written in the thermodynamic limit $N \to \infty $ in the form
$f=\lim_{n\rightarrow 0}=\phi_n/n$ with  $\phi_n$ the replicated free
energy. For the model at hand, assuming at first that
there is only one condensed pattern, say $\mu=1$, we get 
\begin{eqnarray}
&&  \phi_{n,K}=\frac{1}{2}\sum_{a=1}^n m^{\prime 2}_a
     +\frac{9}{2}\alpha '\beta '\sum_{a<b}r^{\prime}_{ab}q^{\prime}_{ab}
           +\frac{9\alpha '}{4\beta '}{\rm Tr}\ln\Lambda
                          \nonumber\\
         &&-\frac{1}{\beta '} \ln
            \left\langle\!\left\langle\sum_{\{s,\sigma\}}
         \exp\left\{\beta '\sum_a m^{\prime}_a b_a
         +\frac{9}{2}\alpha '{\beta '}^2\sum_{a<b}r^{\prime}_{ab} b_{ab}
         \right\}    \right\rangle\!\right\rangle
          \label{free}    \nonumber\\
\end{eqnarray}
where we have dropped the index $1$ and where
\begin{eqnarray}
     b_a&=&s^a\xi+\sigma^a\eta+s^a\sigma^a\xi\eta
            \nonumber\\
  b_{ab}&=&s^as^b+\sigma^a\sigma^b+s^a\sigma^as^b\sigma^b
              \nonumber\\
  \Lambda_{ab}&=&(1-3\beta ')\delta_{ab}-\beta' q^{\prime}_{ab}\, ,
         \quad a,b=1,...,n
               \nonumber \, .
\end{eqnarray}
The brackets $\langle \!\langle \cdots\rangle\!\rangle$
indicate the average over the condensed pattern.
As usual $\beta'$ is the inverse temperature, $\alpha'$ the capacity
defined as the number of patterns per number of couplings per spin, i.e.
$\alpha'=2p/9N$, $\sum_{\{s,\sigma\}}$ denotes the sum over all
configurations at one site and $\sum_{a<b}$ denotes the sum over pairs
of different replicas $a < b$. 
Finally, the set of order parameters is given by
\begin{eqnarray}
   &&m_a^{\prime \mu}=\left\langle\!\left\langle
       \frac{1}{N}\sum_{i=1}^N
           \left\langle s_i^a \right\rangle \xi_i^\mu
           +\left\langle \sigma_i^a \right\rangle \eta_i^\mu
    +\left\langle s_i^a \sigma_i^a \right\rangle \xi_i^\mu \eta_i^\mu
                \right\rangle\!\right\rangle
                \nonumber \\
    &&q'_{ab} =\left\langle\!\left\langle \frac{1}{N}\sum_{i=1}^N
           \left\langle s_i^a\right\rangle
                 \left\langle s_i^b\right\rangle
            + \left\langle \sigma_i^a\right\rangle
                 \left\langle \sigma_i^b\right\rangle
            + \left\langle s_i^a\sigma_i^a\right\rangle
                 \left\langle s_i^b\sigma_i^b\right\rangle
             \right\rangle\!\right\rangle
             \nonumber \\
    &&r'_{ab} =\frac{2}{9 \alpha'}\sum_{\mu >1}^p\left\langle\!\left\langle
              m^{\prime \mu}_a  m^{\prime \mu}_b
             \right\rangle\!\right\rangle
             \nonumber
\end{eqnarray}
where $\langle \cdots \rangle$ denotes the thermal average and the
brackets $\langle \!\langle \cdots\rangle\!\rangle$ now indicate the
average over all patterns.

The order parameters $m'_a$, $q'_{ab}$ and $r'_{ab}$,
and the inverse temperature $\beta '$  can be rescaled in such a way
that the  resulting replicated free energy density (\ref{free}) satisfies
$\phi_{n,K}$ = 3 $\phi_{n,AT}$,  with $\phi_{n,AT}$ the replicated free
energy density for the AT neural network. Hereby we have taken into account 
that for the AT neural network model with equal coupling strengths and
one condensed pattern, the nine order parameters 
($m_{a}^\nu, q_{ab}^\nu, r_{ab}^\nu$) with $\nu=1,2,3$ 
refering to $\xi$, $\eta$ and $\xi\eta$ reduce to  three, i.e.
($m_{a}, q_{ab}, r_{ab}$), where a reference to a specific type of
pattern is now irrelevant. This is due to the fact that for this
AT model only states satisfying 
$m_{a}^\nu =m_{a}, \,  q_{ab}^\nu=q_{ab}, \, r_{ab}^\nu=r_{ab}, \nu=1,2,3$,
i.e. so-called simple states, mimimize the free energy. 
For the replica symmetric ansatz this  property of the simple states has been
shown in \cite{BK2} to be related  
to taking the  quenched average over just one condensed
pattern. Since patterns do not carry replica indices, we 
assume that it is also valid in the fully replicated case.

The proper rescaling is the following~:
\begin{eqnarray}
  &&m'_{a}=3m_{a},~~~q'_{ab}=3q_{ab},~~~r'_{ab}=3r_{ab},\nonumber\\
  &&\beta'=\frac{1}{3}\beta,~~~ \alpha'=\alpha \,.
\end{eqnarray}

Next, assuming more than one condensed pattern,  the order
parameters $\bm_a^\nu$ get a vector character in $\mu$ and stable
states for which the $\bm_a^\nu$ are different for different $\nu$
occur. This no longer allows for a reduction of the order parameters. 
We remark that these states have a bigger replica symmetric free
energy than the one for the simple states and, hence, they play a minor
role in the thermodynamics of the model. Nevertheless, they do destroy the
thermodynamic equivalence with the Potts model.  

In brief, we conclude that the AT neural network with equal
coupling strengths and one condensed pattern is thermodynamically
equivalent  to the four-state
anisotropic Potts model studied in \cite{Kanter}, in spite of the different
Hamiltonians.
In fact, the AT Hamiltonian (\ref{at}) does not contain
three-spin interaction terms present in (\ref{split}).
We have demonstrated this thermodynamic equivalence by rewriting the
Hamiltonian  of
the Potts model using two different types of Ising variables and
calculating the replicated free energy.

These results clarify the resemblance found before \cite{BK1,BK2,BK3} of
the thermodynamic properties of the AT and the Potts neural  
networks. 
Furthermore, they imply that the  four-state Potts model described
by ${\cal H}_{VZ}$ is  thermodynamically equivalent to the AT neural network 
model with one condensed pattern
only in the limit of low loading, i.e. for $\alpha=0$, or at zero
temperature assuming replica symmetry, where we know
that the fixed-point equations for the two four-state Potts models are the
same, as shown in \cite{vz}.

\end{multicols}

\end{document}